\documentclass[a4paper,10pt]{article}
\usepackage{amssymb}
\usepackage{graphicx}

\def\h{\hspace*{5mm}}
\newtheorem{theorem}{Theorem}
\newtheorem{lemma}[theorem]{Lemma}
\newtheorem{corollary}[theorem]{Corollary}
\def\qed{\hfill\hbox{\rlap{$\sqcap$}$\sqcup$}\\\\}
\def\proof{\noindent{\it Proof.\ \ \ }}

\title{Weighted Matching in the Semi-Streaming Model\thanks{Supported by the DFG Research Center {\sc Matheon} ``Mathematics for key technologies" in Berlin}}
\author{Mariano Zelke\thanks{tel.:+49-30-2093 3196, fax:+49-30-2093 3191}
\\ \small Humboldt-Universit\"at zu Berlin,\\\small Institut f\"ur Informatik,\\\small 10099 Berlin\\\small zelke@informatik.hu-berlin.de
}
\date{}

\begin{document}

\maketitle

\begin{abstract}
\noindent We reduce the best known approximation ratio for finding a weighted matching of a graph using a one-pass semi-streaming algorithm from 5.828 to 5.585. The semi-streaming model forbids random access to the input and restricts the memory to ${\cal O}(n\cdot\mbox{polylog}\,n)$ bits. It was introduced by Muthukrishnan in 2003 and is appropriate when dealing with massive graphs.
\end{abstract}

\begin{center}{\small{\bf Keywords:} semi-streaming algorithm, approximation algorithm, matching, graph algorithm}\end{center}

\subsection*{1 Introduction}
\textbf{Matching.} Consider an undirected graph $G=(V,E)$ without multi-edges or loops, where $n$ and $m$ are the number of the vertices and edges, respectively. Let furthermore $w:E\rightarrow \mathbb{R}^{+}$ be a function that assigns a positive weight $w(e)$ to each edge $e$. A \emph{matching} in $G$ is a subset of the edges such that no two edges in the matching have a vertex in common. With $w(M):=\sum_{e\in M}w(e)$ being the weight of a matching $M$, the \emph{maximum weighted matching problem} $MWM$ is to find a matching in $G$ that has maximum weight over all matchings in $G$. 

That problem is well studied and exact solutions in polynomial time are known, see \cite{Schrijver} for an overview. The fastest algorithm is due to Gabow\cite{Gabow1990} and runs in time ${\cal O}(nm+n^2\log n)$.

\noindent\textbf{Approximation Algorithms.} When processing massive graphs even the fastest exact algorithms computing an MWM are too time-consuming. Examples where weighted matchings in massive graphs must be calculated are the refinement of FEM nets \cite{MoehringMueller2000} and multilevel partitioning of graphs \cite{MonienPreisDiekmann2000}. 

To deal with such graphs there has been effort to find algorithms that in a much shorter running time compute solutions that are not necessarily optimal but have some guaranteed quality. Such algorithms are called \emph{approximation algorithms} and their performance is given by an \emph{approximation ratio}. A matching algorithm achieves a $c$-approximation ratio if for all graphs the algorithm finds a matching $M$ such that $w(M)\ge \frac{w(M^*)}{c}$, where $M^*$ is a matching of maximum weight.

A 2-approximation algorithm computing a matching in time ${\cal O}(m)$ was given by Preis \cite{Preis1999}. The best known approximation ratio approachable in linear time is $(3/2+\varepsilon)$ for an arbitrarily small but constant $\varepsilon$. This ratio is obtained by an algorithm of Drake and Hougardy\cite{DrakeHougardy2005} in time ${\cal O}(m\cdot\frac{1}{\varepsilon})$, an algorithm of Pettie and Sanders\cite{PettieSanders04} gets the same ratio slightly faster in time ${\cal O}(m\cdot\log\frac{1}{\varepsilon})$.

\noindent\textbf{Streaming Model. } If we consider graphs being too big to run exact MWM algorithms on them, also an assumption of the classical RAM model is put in question: It is by no means the case that a massive graph can always be assumed as being stored completely within main memory, it is rather stored on disks or even tapes. Now seek times of read/write heads are dominating the running time. Thus for algorithms as the above ones that do not consider the peculiarities of external memory the running time totally get out of hand.

To develop time-efficient algorithms working on these storage devices it is reasonable to assume the input of the algorithm (which is the output of the storage devices) to be a sequential stream. While tapes produce a stream as their natural output, disks reach much higher output rates when presenting their data sequentially in the order it is stored.

Streaming algorithms are developed to deal with such large amounts of data arriving as a stream. In the classical \emph{data stream model}, see e.g. \cite{HenzingerRaghavanRajagopalan99}, \cite{Muthukrishnan}, the algorithm has to process the input stream using a working memory that is small compared to the length of the input. In particular the algorithm is unable to store the whole input and therefore has to make space-efficient summarizations of it according to the query to be answered.

\noindent\textbf{Semi-Streaming Model.} To deal with graph problems in the streaming context Muthukrishnan\cite{Muthukrishnan} proposed the model of a \emph{semi-streaming algorithm}: Random access to the input graph $G$ is forbidden, on the contrary the algorithm gets the edges of $G$ in arbitrary order as the input stream. The memory of the algorithm is restricted to ${\cal O}(n\cdot\mbox{polylog}\,n)$ bits. That does not suffice to store all edges of $G$ if $G$ is sufficiently dense, i.e., $m=\omega(n\cdot\mbox{polylog}\,n)$. A semi-streaming algorithm may read the input stream for a number of $P$ passes. The parameter $T$ denotes the \emph{per-edge processing time}, that is, the time the algorithm needs to handle a single edge.

Despite the heavy restrictions of the model there has been progress in developing semi-streaming algorithms solving graph problems. Feigenbaum et al.\cite{FeigenbaumKannanMcGregorSuriZhang04}, \cite{FeigenbaumKannanMcGregorSuriZhang05} present semi-streaming algorithms for testing $k$-vertex and $k$-edge connectivity of a graph, $k$ being a constant. They point out how to find the connected components and a bipartition and how to calculate a minimum spanning tree of a weighted graph. Zelke\cite{Zelke07} showed how all these problems can be solved using only a constant per-edge processing time.

\noindent\textbf{Matching in the Semi-Streaming Model.} There are approaches to find a weighted matching of a graph in the semi-streaming model. McGregor\cite{McGregor} presents an algorithm finding a $(2+\varepsilon)$-approximative solution with a number of passes $P>1$ depending on $\varepsilon$. 

However, for some real-world applications even a second pass over the input stream is unfeasible. If observed phenomena are not stored and must be processed immediately as they happen only a single pass over the input can occur. For the case of one-pass semi-streaming algorithms it is known, see \cite{FeigenbaumKannanMcGregorSuriZhang04}, that finding the optimal solution to the MWM problem is impossible in general graphs. A first one-pass semi-streaming algorithm approximating the MWM problem with a ratio of 6 presented in \cite{FeigenbaumKannanMcGregorSuriZhang04} was tweaked in \cite{McGregor} to a ratio of 5.828, which was the best known ratio until recently. Both algorithms use only a per-edge processing time of ${\cal O}(1)$.

\noindent\textbf{Our Contribution.} In this paper we present a semi-streaming algorithm that runs in one pass over the input, has a constant per-edge processing time, and that approximates the MWM problem on general graphs with a ratio of 5.585. Therefore it surpasses the known semi-streaming algorithms computing a weighted matching in a single pass. In Section 2 we present our algorithm and its main ideas.  While the proof of the approximation ratio if found in Section 3, we conclude in Section 4.

\subsection*{2 The Algorithm}
In a graph $G=(V,E)$ let two edges be \emph{adjacent} if they have a vertex in common. While $M^*$ denotes a matching of maximum weight in $G$ let in the following $M$ be the matching of $G$ that is currently under consideration by our algorithm. For a set of vertices $W$ we call $M(W)$ to be the set of edges in $M$ covering a vertex in $W$. Correspondingly, for a set $F$ of edges we denote by $M(F)$ all edges in $M$ that are adjacent to an edge in $F$. A set of edges in $E\setminus M$ that are pairwise not adjacent we call an \emph{augmenting set}. Throughout the whole paper $k$ denotes a constant greater than 1.

\begin{figure}
\fbox{\parbox{13.2cm}{
\noindent
{\bfseries Shadow Matching$(G,k)$}\\\ \\
\makebox[10pt][r]{1}
\h $M:=\emptyset$\\
\makebox[10pt][r]{2}
\h {\bfseries while} input stream is not empty\\
\makebox[10pt][r]{3}
\h\h get next input edge $y_1y_2$\\\\
\makebox[10pt][r]{4}
\h\h Let $g_1y_1$, $g_2y_2$ be the edges of $M$ sharing a vertex with $y_1y_2$\\
\makebox[10pt][r]{5}
\h\h $a_1g_1$ $:=$ shadow-edge$(g_1y_1,g_1)$\\
\makebox[10pt][r]{6}
\h\h $a_2g_2$ $:=$ shadow-edge$(g_2y_2,g_2)$\\
\makebox[10pt][r]{7}
\h\h Let $a_1c_1$ be the edge of $M$ covering vertex $a_1$\\
\makebox[10pt][r]{8}
\h\h Let $a_2c_2$ be the edge of $M$ covering vertex $a_2$\\
\makebox[10pt][r]{9}
\h\h $S:=\{y_1y_2,\,g_1y_1,\,a_1g_1,\,a_1c_1,\,g_2y_2,\,a_2g_2,\,a_2c_2\}$\\\\
10
\h\h Find an augmenting set $A\subseteq S$ that maximizes $r(A):=w(A)-k\cdot w(M(A))$\\
11
\h\h {\bfseries if} $r(A)>0$ {\bfseries then}\\ 
12
\h\h\h store each edge in $M(A)$ as a shadow-edge of its adjacent edges in $A$\\
13
\h\h\h $M:=(M\setminus M(A))\cup A$
}}
\caption{\label{MainLoop}The algorithm Shadow Matching}

\end{figure}
\begin{figure}[b]
\begin{center}
\includegraphics[width=220pt]{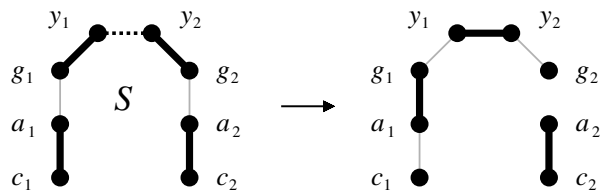}
\parbox{351pt}{\caption{\label{PictureAlgStep}Example of an algorithm's step. Edges in $M$ are shown in bold, shadow-edges appear in grey. $y_1y_2$ is the actual input edge shown dashed. The algorithm inserts the augmenting set $A=\{y_1y_2, a_1g_1\}$ into $M$. Therefore the edges $M(A)=\{a_1c_1,g_1y_1,g_2y_2\}$ are removed from $M$, they become shadow-edges.}}
\end{center}
\end{figure}

Our algorithm is given in Figure \ref{MainLoop}. Note at first that each edge in the algorithm is denoted by its endpoints, which is done for the sake of simpler considerations in the following on edges having common vertices. Every edge is well-defined by its endpoints since we assume the input graph $G$ to contain neither multi-edges nor loops.

The general idea of the algorithm is to keep a matching $M$ of $G$ at all times and to decide for each incoming edge $y_1y_2$ in the input stream if it is inserted into $M$. This is the case if the weight of $y_1y_2$ is big compared to the edges already in $M$ sharing a vertex with $y_1y_2$ and that therefore must be removed from $M$ to incorporate $y_1y_2$. 

This idea so far has already been utilized by one-pass semi-streaming algorithms of Feigenbaum et al.\cite{FeigenbaumKannanMcGregorSuriZhang04} and McGregor\cite{McGregor} seeking a matching in weighted graphs. However, our algorithm differs from the ones in \cite{FeigenbaumKannanMcGregorSuriZhang04} and \cite{McGregor} in fundamental points.

First, if the algorithms in \cite{FeigenbaumKannanMcGregorSuriZhang04} and \cite{McGregor} remove an edge from the actual matching $M$ this is irrevocable. Our new algorithm, by contrast, stores some edges that have been in $M$ in the past but were removed from it. To potentially reinsert them into $M$ the algorithm memorizes such edges under the name of shadow-edges. For an edge $xy$ in $M$ \emph{shadow-edge$(xy,a)$},  $a\in\{x,y\}$, denotes an edge that is stored by the algorithm and shares the vertex $a$ with $xy$. Every edge $xy$ in $M$ has at most two shadow-edges assigned to it, at most one shadow-edge is assigned to the endpoint $x$ and at most one is assigned to $y$.

A second main difference is the way of deciding if an edge $e$ is inserted into $M$ or not. In the algorithms of \cite{FeigenbaumKannanMcGregorSuriZhang04} and \cite{McGregor} this decision is based only on the edges in $M$ adjacent to $e$. Our algorithm takes edges in $M$ as well as shadow-edges in the vicinity of $e$ into account to decide the insertion of $e$.

Finally the algorithms of \cite{FeigenbaumKannanMcGregorSuriZhang04} and \cite{McGregor} are limited to the inclusion of the actual input edge into $M$. By reintegrating shadow-edges our algorithm can insert up to three edges into $M$ within a single step.

\ \\
Let us take a closer look at the algorithm. As an example of a step of the algorithm, Figure \ref{PictureAlgStep} is given. But note that this picture shows only one possible configuration of the set $S$. Since non-matching edges in $S$ may be adjacent, $S$ may look different.

After reading the actual input edge $y_1y_2$ the algorithm tags all memorized edges in the vicinity of $y_1y_2$. This is done in lines 4-8. If an edge is not present the corresponding tag denotes the null-edge, that is, the empty set of weight zero. Thus if for example the endpoint $y_2$ of the input edge $y_1y_2$ is not covered by an edge in $M$, the identifier $g_2y_2$ denotes a null-edge, as well as its shadow-edge $a_2g_2$ and the edge $a_2c_2$. All edges tagged so far are taken into consideration in the remaining part of the loop, they are subsumed to the set $S$ in line 9. 

In line 10 all augmenting sets of $S$ are examined. Among these sets the algorithm selects $A$ that maximizes $r(A)$. If $r(A)>0$ the edges of $A$ are taken into $M$ and the edges in $M$ sharing a vertex with edges in $A$ are removed from $M$. We say $A$ is inserted into $M$, this is done in line 13.

If an augmenting set $A$ is inserted into $M$ this is always accompanied by storing the removed edges $M(A)$ as shadow-edges of edges in $A$ in line 12. More precisely, every edge $e$ in $M(A)$ is assigned as a shadow-edge to every edge in $A$ that shares a vertex with $e$. If, as in the example given in Figure \ref{PictureAlgStep}, $A=\{y_1y_2, a_1g_1\}$, the edge $g_1y_1$ that is adjacent to both edges in $A$ is memorized under the name shadow-edge$(y_1y_2,y_1)$ as well as under the name shadow-edge$(a_1g_1,g_1)$. $a_1c_1$ is stored as shadow-edge$(a_1g_1,a_1)$, $g_2y_2$ as shadow-edge$(y_1y_2,y_2)$. After inserting $A$, $a_2g_2$ is not memorized as a shadow-edge assigned to $g_2y_2$ since $g_2y_2$ is not an edge in $M$ afterwards. That is indicated in Figure \ref{PictureAlgStep} by the disappearance of $a_2g_2$. However, if $a_2g_2$ was memorized as a shadow-edge of $a_2c_2$ before, this will also be the case after inserting $A$.

It is important to note that there is never an edge in $M$ which is a shadow-edge at the same time: Edges only become shadow-edges if they are removed from $M$. An edge which is inserted into $M$ is no shadow-edge anymore, since there is no edge in $M$ it could be assigned to as a shadow-edge. 

\ \\
It is easy to see that our algorithm computes a valid matching of the input graph $G$.

\begin{corollary}\label{validMatching}
Throughout the algorithm Shadow Matching($G,k$), $M$ is a matching of $G$.
\end{corollary}
\proof This is true at the beginning of the algorithm since $M=\emptyset$. Whenever the algorithm modifies $M$ in line 13 it inserts the edges in $A$, they are pairwise not adjacent, and removes all edges $M(A)$ that are adjacent to an edge in $A$. Thus $M$ never includes two adjacent edges. \qed 

\noindent Our algorithm may remind of algorithms in \cite{DrakeHougardy2005} and \cite{PettieSanders04} approximating a maximum weighted matching in the RAM model. Starting from some actual matching $M$ in a graph $G$ these algorithms look for short augmentations, that is, connected subgraphs of $G$ having constant size in which edges in $M$ and $E\setminus M$ can be exchanged to increase the weight of the actual matching. 

From this point of view our algorithm may suggest itself as it is reasonable to expect the notion of short augmentations to be profitable in the semi-streaming model as well. However, we are unable to use even the basic ideas of proving the approximation ratio in \cite{DrakeHougardy2005} and \cite{PettieSanders04}. As well as the algorithms the proof concept relies on random access to the whole graph, a potential we cannot count on in the semi-streaming model.

Certainly our algorithm can be considered as a natural extension of the semi-streaming algorithms in \cite{FeigenbaumKannanMcGregorSuriZhang04} and \cite{McGregor} seeking a weighted matching. But the abilities of our algorithm go beyond the insertion of a single edge to the actual matching, the step to which the algorithms in \cite{FeigenbaumKannanMcGregorSuriZhang04} and \cite{McGregor} are limited to. Therefore we have to substantially enhance the proof techniques used therein to attest an improved approximation ratio of our algorithm. This is done in the next section.

\subsection*{3 Approximation Ratio}
Consider an augmenting set $A$ which covers the vertices $B$ and let $k>1$ be some constant. We call $f_{A,k}:V\rightarrow \{x\in\mathbb{R}\,|\,0\le x\le 1\}$ an \emph{allocation function} for $A$ if $f_{A,k}(v) = 0$ for all $v\in V\setminus B$ and additionally the following holds:
\begin{itemize}
 \item $\forall$ $ab\in A:f_{A,k}(a)\cdot w(M(a))+f_{A,k}(b)\cdot w(M(b))\le \frac{w(ab)}{k}$
 \item $\forall$ $cd\in M(A): f_{A,k}(c)+f_{A,k}(d)\ge 1$
\end{itemize}
If there exists such an allocation function $f_{A,k}$ for an augmenting set $A$ we call $A$ to be \emph{locally $k$-exceeding}.
The intuition here is as follows: If for an augmenting set $A$ we have $w(A)>k\cdot w(M(A))$ we can distribute the weight of the edges in $M(A)$ to the edges of $A$ in such a way that every edge $ab$ in $A$ gets weight of at most $\frac{w(ab)}{k}$ distributed to it. If $A$ satisfies the stronger condition of being locally $k$-exceeding such a weight distribution can also be done with the additional property that the weight of an edge $cd$ in $M(A)$ is distributed only to edges in $A$ that are adjacent to $cd$.

\begin{lemma}\label{AisLocallyKExceeding}
Every augmenting set $A$ that is inserted into $M$ by the algorithm Shadow Matching$(G,k)$ is locally $k$-exceeding.
\end{lemma}
\proof Since $A\subseteq\{y_1y_2,a_1g_1,a_2g_2\}$ and $r(A)>0$, $1\le|A|\le3$. If $A$ consists of only one edge, say $y_1y_2$, we have for the sum of the weights of the adjacent edges $w(g_1y_1)+w(g_2y_2)\le \frac{w(y_1y_2)}{k}$ because of the satisfied condition in line 11. In that case the allocation function is $f_{A,k}(y_1)=f_{A,k}(y_2)=1$ and $A$ is locally $k$-exceeding.

Let $A$ consist of two edges, say $y_1y_2$ and $a_1g_1$. Since every subset of $A$ is an augmenting set as well which is not taken by the algorithm we know that $r(\{y_1y_2,a_1g_1\})\ge r(\{y_1y_2\})$ and therefore
$$w(y_1y_2)+w(a_1g_1)-k(w(a_1c_1)+w(g_1y_1)+w(g_2y_2))\ge w(y_1y_2)-k(w(g_1y_1)+w(g_2y_2))$$
Thus $w(a_1g_1)\ge k\cdot w(a_1c_1)$ and because $r(\{y_1y_2,a_1g_1\})\ge r(\{a_1g_1\})$ we can deduce similarly $w(y_1y_2)\ge k\cdot w(g_2y_2)$. Hence for the allocation function we can set $f_{A,k}(a_1)=f_{A,k}(y_2)=1$. Since $r(A)>0$ we can find appropriate values for $f_{A,k}(g_1)$ and $f_{A,k}(y_1)$, therefore $A$ is locally $k$-exceeding.

For other configurations of $A$ it can be exploited correspondingly that $r(A)\ge r(A^\prime)$ for all subsets $A^\prime$ of $A$. Therefore it can be shown similarly that an allocation function exists and $A$ is locally $k$-exceeding.\qed 

\noindent Because of Corollary \ref{validMatching} we can take the final $M$ of the algorithm as a valid solution for the weighted matching problem on the input graph $G$. It is immediate that the constant $k$ is crucial for the weight of the solution we get and therefore determines the ratio up to which the algorithm approximates an optimal matching. The main part of the paper is to prove the following theorem which we just state here and which we prove later.

\begin{theorem}\label{ApproxRatio}
The algorithm Shadow Matching($G,k$), $k>1$, constructs a weighted matching $M$ of $G$. $M$ is at most a factor of $$k+\frac{k}{k-1}+\frac{k^3-k+1}{k^2}$$ smaller than the weight of an optimal weighted matching of $G$.
\end{theorem}

\noindent We call $G_i$ the subgraph of $G$ consisting of the first $i$ input edges, $M_i$ denotes the $M$ of the algorithm after completing the while-loop for the $i$th input edge. An edge $xy$ \emph{prevents} an edge $ab$ if $ab$ is the $i$th input edge and $xy\in M_i$ shares an endpoint with $ab$, thus $ab$ is not taken into $M$ by the algorithm. Note that an edge might be prevented by one or two edges. An edge $xy$ \emph{replaces} an edge $cd$ if $xy$ is the $i$th input edge, $xy$ and $cd$ share a vertex, $cd\in M_{i-1}$, and $xy\in M_i$. Therefore $cd$ is not in $M$ afterwards. An edge can replace up to two edges and can be replaced by up to two edges.

Consider an optimal solution $M^*=\{o_1,o_2,\ldots\}$ for the MWM problem of $G$, $M^*_i:=M^*\cap G_i$. The edges $o_1,o_2,\ldots$ in $M^*$ we call \emph{optimal edges}. If $w(M_i)<w(M^*_i)$, some edges of $M^*_i$ must be missing in $M_i$. There are two possible reasons for the absence of an edge $o_l\in M^*_i$ in $M_i$. First, there are edges in $M_j$, $j<i$, which prevented $o_l$. Second, $o_l\in M_j$, $j<i$, is replaced by one or more edges and never reinserted.

In any case we can make edges in $\bigcup_{h\le i} M_h$ responsible for missing edges of $M^*_i$ in $M_i$. We charge the weight of an optimal edge $o_l$ to the edges in $\bigcup_{h\le i} M_h$ that are responsible for the prevention or the removal of $o_l$. If such a charged edge in $M$ is replaced by other edges its charge is transferred to the replacing edges such that no charge is lost. After all we can sum up the charges of all edges in the final $M_m$ to get $w(M^*\setminus M_m)$.

To bound $w(M^*_i\setminus M_i)$ as a multiple $c$ of $w(M_i)$ if suffices to show that each edge $xy\in M_i$ carries a charge of at most $c\cdot w(xy)$. This technique has been carried out by Feigenbaum et al.\cite{FeigenbaumKannanMcGregorSuriZhang04} and McGregor\cite{McGregor} to estimate the approximation ratios of their semi-streaming algorithms calculating a weighted matching.

We follow the same general idea but need a more sophisticated approach of managing the charge. This is due to two reasons. First, the algorithms of \cite{FeigenbaumKannanMcGregorSuriZhang04} and \cite{McGregor} are limited to a simple replacement step which substitutes one or two edges by a single edge $e$. That makes the charge transfer easy to follow since the charges of the substituted edges are transferred completely to the single edge $e$. Our algorithm, by contrast, is able to substitute several edges by groups of edges. The charge to be transferred must be distributed carefully to the replacing edges. 

Second, in the algorithms of \cite{FeigenbaumKannanMcGregorSuriZhang04} and \cite{McGregor} the decision whether to insert an input edge into $M$ is determined only by the edges in $M$ adjacent to the input edge. If an optimal edge $o$ is not taken into $M$ the charge can simply be assigned to the at most two edges already in $M$ that are adjacent to $o$. In our algorithm not only the edges in $M$ that are adjacent to $o$ specify if $o$ is taken into $M$. In fact, several shadow-edges and other edges in $M$ in the environment of $o$ may codetermine if $o$ is inserted into $M$. These ambient edges must be taken into account if charge has to be distributed for preventing $o$. 

For our more sophisticated technique of managing the charges we think of every edge $xy\in M$ as being equipped with two values, namely \emph{charge of optimal edge coe}$(xy,x)$ and $coe(xy,y)$, one for every endpoint of $xy$. $coe(xy,x)$ is the charge that the edge in $M^*$ which is covering the vertex $x$ is charging to $xy$.

If an edge is removed from $M$ its charges are transfered to the one or two replacing edges. Therefore in addition to its $coe(xy,x)$ and $coe(xy,y)$ every edge $xy\in M$ is equipped with a third value \emph{aggregated charge} $ac(xy)$ which is the sum of the charges that $xy$ takes over from edges replaced by $xy$. We define $T(xy):=coe(xy,x)+coe(xy,y)+ac(xy)$ as the sum of the charges of the edge $xy$.

During the proof of the following lemma we will explicitly show how the weights of edges in $M^*_i\setminus M_i$ can be charged to the edges in $M_i$ and how these charges are transferred to replacing edges such that particular properties hold. 

\begin{lemma}\label{Messy}
Let $M_i$ be the solution found by the algorithm Shadow Matching($G,k$), $k>1$, after reading $G_i$ for $1\le i\le m$. To every edge $xy$ in $M_i$ we can assign three values $coe(xy,x)$, $coe(xy,y)$ and $ac(xy)$, with $T(xy)$ being their sum, such that:
\begin{itemize}
\item[a)]$\sum\limits_{xy\in M_i}T(xy)\ge w(M^*_i\setminus M_i)$
\item[b)]$\forall$ $xy\in M_i$: $\,coe(xy,x)\le k\cdot w(xy)$ and  $coe(xy,y)\le k\cdot w(xy)$ 
\item[c)]$\forall$ $xy\in M_i$: $\,ac(xy)\le\frac{k}{k-1}\cdot w(xy)$
\item[d)]$\forall$ $xy\in M_i$: $\,T(xy)\le\left(k+\frac{k}{k-1}+\frac{k^3-k+1}{k^2}\right)\cdot w(xy)$
\end{itemize}
\end{lemma}

\proof Let $y_1y_2$ be the actual input edge. We first take a look at the different cases that can occur if $y_1y_2$ is not taken into $M$ by the algorithm. We postpone the cases in which $a_1g_1$ and $a_2g_2$ are adjacent, thus until further notice the set $\{a_1g_1,a_2g_2\}$ is an augmenting set. If $y_1y_2$ as the actual input edge is not taken into $M$ this is due to two possible reasons. First, no augmenting set is inserted into $M$. Second, the augmenting set that is inserted does not contain $y_1y_2$. If the first case occurs at least one of the following conditions is satisfied:
\begin{itemize}
\item[i)] $w(y_1y_2)\le\min\{k\cdot(w(g_1y_1)+w(g_2y_2))$,\\
$k\cdot(w(g_1y_1)+w(a_1c_1))-w(a_1g_1)+k\cdot w(g_2y_2)$,\\
$k\cdot(w(g_2y_2)+w(a_2c_2))-w(a_2g_2)+k\cdot w(g_1y_1)$,\\ $k\cdot(w(g_1y_1)+w(a_1c_1))-w(a_1g_1)+k\cdot(w(g_2y_2)+w(a_2c_2))-w(a_2g_2)\}$

\item[ii)] $w(y_1y_2)\le\min\{k\cdot(w(g_1y_1)+w(g_2y_2))$,\\
$k\cdot(w(g_2y_2)+w(a_2c_2))-w(a_2g_2)+k\cdot w(g_1y_1)\}$ and $y_2=a_1$

\item[iii)] $w(y_1y_2)\le\min\{k\cdot(w(g_1y_1)+w(g_2y_2))$,\\
$k\cdot(w(g_1y_1)+w(a_1c_1))-w(a_1g_1)+k\cdot w(g_2y_2)\}$ and $y_1=a_2$
\item[iv)] $w(y_1y_2)\le k\cdot(w(g_1y_1)+w(g_2y_2))$, $y_1=a_2$, and $y_2=a_1$
\end{itemize}
Condition $i)$ captures the situation where $\{y_1y_2,a_1g_1,a_2g_2\}$ is an augmenting set but neither this set nor one of its subsets satisfies the condition in line 11 of the algorithm. Conditions $ii)-iv)$ include the cases in which $y_1y_2$ is adjacent to $a_1g_1$, $a_2g_2$, or both.

As mentioned above the algorithm possibly inserts an augmenting set $A$ into $M$ that does not contain $y_1y_2$. Exploiting the fact that $r(A)\ge r(A^\prime)$ for all other augmenting sets $A^\prime$ we get that at least one of the following conditions is satisfied in this case.
\begin{itemize}
\item[v)] $w(y_1y_2)\le\min\{k\cdot w(g_2y_2)$, $k\cdot(w(g_2y_2)+w(a_2c_2))-w(a_2g_2)\}$ and $A=\{a_1g_1\}$
\item[vi)] $w(y_1y_2)\le\min\{k\cdot w(g_1y_1)$, $k\cdot(w(g_1y_1)+w(a_1c_1))-w(a_1g_1)\}$ and $A=\{a_2g_2\}$
\item[vii)] $w(y_1y_2)\le k\cdot w(g_2y_2)$, $y_1=a_2$, and $A=\{a_1g_1\}$
\item[viii)] $w(y_1y_2)\le k\cdot w(g_1y_1)$, $y_2=a_1$, and $A=\{a_2g_2\}$
\item[ix)] $w(y_1y_2)\le w(a_1g_1)$, $y_2=a_1$, and $\{a_1g_1\}\subseteq A$
\item[x)] $w(y_1y_2)\le w(a_2g_2)$, $y_1=a_2$, and $\{a_2g_2\}\subseteq A$
\item[xi)] $w(y_1y_2)\le w(a_1g_1)+w(a_2g_2)$, $y_2=a_1$, $y_1=a_2$, and $A=\{a_1g_1,a_2g_2\}$
\end{itemize}
If the $i$th input edge $y_1y_2\in M^*$ is not taken into $M_i$ we have to charge the edges in $M_i$ that prevent the optimal edge $y_1y_2$. In the cases $i)-viii)$ the edges $g_1y_1$ and/or $g_2y_2$ prevent $y_1y_2$, in the cases $ix)-xi)$ $a_1g_1$ and/or $a_2g_2$ prevent $y_1y_2$. To charge the preventing edges in $M_i$ we split $w(y_1y_2)$ into two partial weights and charge one partial weight to the edge in $M_i$ covering $y_1$ and one to the edge in $M_i$ covering $y_2$. In any of the above cases $w(y_1y_2)$ can be split into two partial weights in such a way that the following generalization holds.

Let $ab\in M_i$ share the vertex $a$ with the $i$th input edge $o\in M^*$. Let $bc$ be the shadow-edge$(ab,b)$, that is, the shadow-edge assigned to the vertex of $ab$ that is not shared by $o$. Let $cd$ be the edge in $M_i$ that covers $c$. $w(o)$ can be split into two partial weights such that for the partial weight $p$ that $ab$ has to take as a charge for preventing $o$ at least one of the following conditions is satisfied:
\begin{itemize}
\item[(A)] $p\le k\cdot w(ab) \le k\cdot(w(ab)+w(cd))-w(bc)$

\item[(B)] $p\le k\cdot(w(ab)+w(cd))-w(bc) \le k\cdot w(ab)$

\item[(C)] $p\le k\cdot w(ab)$ and $ab$, input edge $o$ and shadow-edge $bc$ form a triangle.
\end{itemize}
We start to prove the lemma by induction over the edges inserted into $M$. More precisely we suppose that the edge $y_1y_2$ as the $i$th input edge is inserted into $M_{i-1}$ and that before this insertion all properties of the lemma are satisfied. 

We have to consider two things: First, we have to point out how the charges of the edges in $M_{i-1}$ that $y_1y_2$ replaces are carried over to $y_1y_2$ to preserve the properties of the lemma. Second we have to regard the at most two optimal edges that possibly come after $y_1y_2$ and share a vertex with $y_1y_2$. If $y_1y_2$ prevents one or both of these edges we have to show how $y_1y_2$ is charged by them without violating the lemma.\\

\noindent For the initial step of our induction note that the properties of the lemma hold for the first input edge.

For the inductive step let $y_1y_2$ as the $i$th input edge be taken into $M_i$. Thus $y_1y_2$ is contained in the augmenting set $A$ that is inserted into $M$. Because of Lemma \ref{AisLocallyKExceeding} $A$ is locally $k$-exceeding, hence there exists an allocation function $f_{A,k}$.

Let in the following $x\in\{1,2\}$. $y_1y_2$ takes over charges from $g_xy_x$, the edges it replaces, according to the allocation function $f_{A,k}$. More precisely it takes over a $f_{A,k}(y_x)$-fraction of the charges of $g_xy_x$. In fact, $y_1y_2$ builds its $ac$ as follows: $ac(y_1y_2)=(coe(g_1y_1,g_1)+ac(g_1y_1))\cdot f_{A,k}(y_1)+(coe(g_2y_2,g_2)+ac(g_2y_2))\cdot f_{A,k}(y_2)$. By the induction hypothesis $coe(g_xy_x,g_x)\le k\cdot w(g_xy_x)$ and $ac(g_xy_x)\le \frac{k}{k-1}\cdot w(g_xy_x)$. Due to the definition of an allocation function $f_{A,k}(y_1)\cdot w(g_1y_1)+f_{A,k}(y_2)\cdot w(g_2y_2)\le \frac{w(y_1y_2)}{k}$. Thus $ac(y_1y_2)\le \frac{k}{k-1}\cdot w(y_1y_2)$ satisfying property c).

Furthermore $y_1y_2$ takes over charge from $coe(g_xy_x,y_x)$ to its own $coe(y_1y_2,y_x)$, again a $f_{A,k}(y_x)$-fraction of it. If $g_xy_x$ is in $M^*$, $coe(g_xy_x,y_x)=0$ and $y_1y_2$ instead takes over a $f_{A,k}(y_x)$-fraction of $w(g_xy_x)$ as its $coe(y_1y_2,y_x)$ for replacing the optimal edge $g_xy_x$.

Note that whenever $f_{A,k}(y_x)<1$, $y_1y_2$ does not take over all the charge of $g_xy_x$. However, the definition of the allocation function makes sure that $f_{A,k}(g_x)\ge 1-f_{A,k}(y_x)$ and that another edge in $A$ covering $g_x$ takes over the remaining charge of $g_xy_x$. That way no charge can get lost and property a) holds.

Let us check the validity of property b). Right after $y_1y_2$ was inserted into $M$ and took over the charges as described from $g_xy_x$ it holds that $coe(y_1y_2,y_x)\le w(y_1y_2)$. That does not suffice to show validity of property b). In fact, there might be an optimal edge $o_xy_x$ coming after $y_1y_2$ in the input stream covering $y_x$. In that case $coe(y_1y_2,y_x)=0$ up to this moment, since there cannot be another optimal edge besides $o_xy_x$ covering $y_x$. If $o_xy_x$ is not inserted into $M$, that is, $y_1y_2$ prevents $o_xy_x$, $y_1y_2$ must be charged. By the considerations above we know about the charges that an edge in $M$ has to take because of optimal edges prevented by it. In all three possibilities (A)-(C) the charge $y_1y_2$ has to include into $coe(y_1y_2,y_x)$ for preventing $o_xy_x$ is at most $k\cdot w(y_1y_2)$, satisfying property b).

It remains to show that property d) holds which bounds the sum of all charges of $y_1y_2$. The situation is as follows: $y_1y_2$ is in $M$ and we call the shadow-edge$(y_1y_2,y_1)$ $g_1y_1$, the shadow-edge$(y_1y_2,y_2)$ $g_2y_2$. Note again that $y_1y_2$ took over only a $f_{A,k}(y_x)$-fraction of the charges from $g_xy_x$. Directly after $y_1y_2$ was inserted into $M$ and took over the charges from the replaced edges as described property d) holds. We have to consider optimal edges $o_xy_x$ that appear after $y_1y_2$ in the input stream, are prevented by $y_1y_2$ and therefore cause charge $p_x$ at $coe(y_1y_2,y_x)$. 

As described $ac(y_1y_2)$ is composed of four values, namely fractions of $ac(g_xy_x)$ and $coe(g_xy_y,g_x)$. The value of the fraction of $ac(g_xy_x)$ that is part of $ac(y_1y_2)$ we call $ac(g_xy_x)\curvearrowright ac(y_1y_2)$, correspondingly we have  $coe(g_xy_x,g_x)\curvearrowright ac(y_1y_2)$. Using that we can separate $T(y_1y_2)$ into two halves as follows
\begin{eqnarray*}
T(y_1y_2)&=&\Big(coe(y_1y_2,y_2)+ac(g_1y_1)\curvearrowright ac(y_1y_2) +coe(g_1y_1,g_1)\curvearrowright ac(y_1y_2)\Big)+\\
&&\Big(coe(y_1y_2,y_1)+ac(g_2y_2)\curvearrowright ac(y_1y_2)+coe(g_2y_2,g_2)\curvearrowright ac(y_1y_2)\Big)
\end{eqnarray*}
Let us call the upper half $H1$ and the lower one $H2$. We will estimate $H2$ in the following according to the three possible cases for $p_1$ and show that 
\begin{eqnarray}
H2 \le \left(k+\frac{1}{k-1}+\frac{1}{k} \right)w(g_2y_2)\cdot f_{A,k}(y_2)+k\cdot w(y_1y_2)
\end{eqnarray}
We will see later that it suffices to show that if neither $H2$ violates (1) nor $H1$ violates a corresponding inequality, property d) holds for $y_1y_2$.\\

\noindent{\it Charge $p_1$ coming from $o_1y_1$ satisfies }(A)\\
Let $g_2z_2$ be an edge in $M$ covering $g_2$. We can bound $p_1$ because of property (A) 
\begin{equation}
p_1\le k\cdot w(y_1y_2)\le k\cdot(w(y_1y_2)+w(g_2z_2))-w(g_2y_2)
\end{equation}
We call the shadow-edge  $g_2y_2$ of $y_1y_2$ \emph{overloaded} if  we have $coe(g_2y_2,g_2)\curvearrowright ac(y_1y_2)>w(g_2y_2)\cdot f_{A,k}(y_2)$. For a shadow-edge $uv$ we say that $uv$ \emph{fingers} $v$ if $uv$ covers $v$ and $v$ is not the vertex that $uv$ shares with the edge in $M$ it is assigned to. For example the shadow-edge $g_2y_2$, which is assigned to $y_1y_2$, fingers $g_2$ but not $y_2$. A shadow-edge $uv$ is \emph{prepared} if for the edge $uw$ in $M$ that $uv$ is assigned to $coe(uw,w)=0$. So in the present example $g_2y_2$ is prepared if $coe(y_1y_2,y_1)=0$. 

If $p_1\le k\cdot w(y_1y_2)-f_{A,k}(y_2)\cdot w(g_2y_2)$ or if $g_2y_2$ is not overloaded, we can simply add $p_1$ to $coe(y_1y_2,y_1)$ and $H2$ satisfies (1). Otherwise we do a \emph{charge transfer} as follows: We reduce $coe(g_2y_2,g_2)\curvearrowright ac(y_1y_2)$ to $r:=\max\{coe(g_2y_2,g_2)\curvearrowright ac(y_1y_2)-(k-1)\cdot w(g_2z_2),0\}$ and add a value of $coe(g_2y_2,g_2)\curvearrowright ac(y_1y_2)-r$ to $coe(g_2z_2,g_2)$, thus no charge is lost.

It is important to see that this increasing of $coe(g_2z_2,g_2)$ does not violate the properties of the lemma for $g_2z_2$: We know that $coe(g_2z_2,z_2)\le k\cdot w(g_2z_2)$ and $ac(g_2z_2)\le \frac{k}{k-1}\cdot w(g_2z_2)$. If before the charge transfer $coe(g_2z_2,g_2)=0$, after the transfer $T(g_2z_2)$ cannot exceed $(k+\frac{k}{k-1}+\frac{k^3-k+1}{k^2})\cdot w(g_2z_2)$.

For the other case, i.e., that $coe(g_2z_2,g_2)>0$ before the charge transfer we need a few considerations. In fact, we will show that for every vertex $v$ at every moment of the algorithm at most one shadow-edge fingers $v$, is overloaded, and prepared at the same time:

Assume that $uv$ is the first shadow-edge created by the algorithm that is fingering $v$ and that is overloaded and prepared. This can only be the case if $uv$ in $M$ gets replaced by $uw$ and possibly $vs$. $uv$ as a shadow-edge of $uw$ is now fingering $v$ and it is overloaded and prepared. Right after the replacement $coe(vs,v)\le w(vs)$. As long as no charge of $coe(uv,v)\curvearrowright ac(uw)$ is transferred to an edge in $M$ covering $v$, for every edge $vq$ in $M$ $coe(vq,v)\le w(vq)$. Such an edge $vq$ cannot be turned into a shadow-edge fingering $v$ and being overloaded. A second overloaded shadow-edge fingering $v$ can only be created by replacing an edge $vr$ with $coe(vr,v)> w(vr)$, that can only occur if $uw$ transfers charge to $vr$. However, $uw$ only transfers charge to $vr$ if it prevents an optimal edge. After that $coe(uw,w)>0$ and $uv$ is not prepared anymore. This shows that a prepared and overloaded shadow-edge fingering $v$ can only be created if the at most one previously prepared and overloaded shadow-edge fingering $v$ lost its status as being prepared. 

Now we can come back to the case $coe(g_2z_2,g_2)>0$. We can assume that $g_2z_2$ as part of the augmenting set $A^\prime$ replaced the edges $d_2g_2$ and $t_2z_2$. $g_2z_2$ took over a $f_{A^\prime,k}(g_2)$-fraction of the charges from $d_2g_2$. Since $coe(d_2g_2,g_2)\le k\cdot w(d_2g_2)$ before the replacement of $d_2g_2$, we have $coe(g_2z_2,g_2)\le f_{A^\prime,k}(g_2)\cdot k\cdot w(d_2g_2)$ after the replacement. By the definition of an allocation function it follows $coe(g_2z_2,g_2)\le w(g_2z_2)-f_{A^\prime,k}(z_2)\cdot k\cdot w(t_2z_2)$. After our charge transfer of weight at most $(k-1)\cdot w(g_2z_2)$ from $coe(g_2y_2,g_2)\curvearrowright ac(y_1y_2)$ to $coe(g_2z_2,g_2)$, it holds that $coe(g_2z_2,g_2)\le k\cdot(w(g_2z_2)-f_{A^\prime,k}(z_2)\cdot w(t_2z_2))$. Therefore the charges of $g_2z_2$ satisfy an inequality corresponding to (1), thus property d) cannot be violated for $g_2z_2$. 

Now the above considerations are important: We know that no shadow-edge besides $g_2y_2$ that is fingering $g_2$ is prepared and overloaded. Thus no further charge transfer to $coe(g_2z_2,g_2)$ can occur violating the properties of the lemma for $g_2z_2$.

After transferring a part of $coe(g_2y_2,g_2)\curvearrowright ac(y_1y_2)$  as described we have $coe(g_2y_2,g_2)$ $\curvearrowright ac(y_1y_2)$ $\le \max\{k\cdot f_{A,k}(y_2)\cdot w(g_2y_2)-(k-1)\cdot w(g_2z_2),0\}$. We add $p_1$ to $coe(y_1y_2,y_1)$ and can evaluate $H2$: We have $coe(y_1y_2,y_1)=p_1\le k\cdot w(y_1y_2)$ because of (2) and $ac(g_2y_2)\curvearrowright ac(y_1y_2)\le f_{A,k}(y_2)\cdot w(g_2y_2)\cdot \frac{k}{k-1}$ by the induction hypothesis. Since $w(g_2z_2)\ge \frac{w(g_2y_2)}{k}$ because of (2) we can estimate $H2$ as being bounded as in (1).\\

\noindent{\it Charge $p_1$ coming from $o_1y_1$ satisfies }(B)\\
This case is very similar to the previous one with the only difference that $w(g_2z_2)\le \frac{w(g_2y_2)}{k}$ and we use $p_1\le k\cdot(w(y_1y_2)+w(g_2z_2))-w(g_2y_2)$. All other considerations remain the same and that results in the very same estimation for $H2$.
\\

\noindent{\it Charge $p_1$ coming from $o_1y_1$ satisfies }(C)\\
In this case $o_1=g_2$ since the input edge $o_1y_1$, the edge $y_1y_2\in M$ and the shadow-edge $g_2y_2$ form a triangle. Since $g_2y_1$ is an optimal edge, before its arrival $coe(g_2y_2,g_2)\curvearrowright ac(y_1y_2)=0$. So $y_1y_2$ can take a charge of $p_1\le k\cdot w(y_1y_2)$ as its $coe(y_1y_2,y_1)$ and $H2$ satisfies (1).\\

\noindent We can handle the charge $p_1$ in every possible case such that $H2$ satisfies (1). With a symmetric argumentation we can show that $H1$ satisfies a corresponding inequality. Using that $f_{A,k}(y_1)\cdot w(g_1y_1)+f_{A,k}(y_2)\cdot w(g_2y_2)\le \frac{w(y_1y_2)}{k}$ we get
\begin{eqnarray*}
T(y_1y_2)=H1+H2\le\left(k+\frac{k}{k-1}+\frac{k^3-k+1}{k^2}\right)\cdot w(y_1y_2)
\end{eqnarray*}\ \\

\noindent It remains to consider the postponed cases in which $y_1y_2$ is not inserted into $M$ by the algorithm and $a_1g_1$, $a_2g_2$ have a vertex in common, hence cannot be taken into $M$ simultaneously. 

If $a_1=g_2$, the augmenting set $A=\{y_1y_2,a_1g_1\}$ and $M(A)$ build a cycle on 4 vertices. If on the one hand $a_2\not=y_1$ in this situation and no augmenting set is inserted into $M$, case $i)$ is satisfied, if $a_2g_2$ is inserted into $M$ case $vi)$ is met. If on the other hand $a_2=y_1$ the cases $iii)$ or $x)$ are applicable.

If $a_1=a_2$ and $a_2g_2$ is taken into $M$, after that $a_2g_2$ is the edge in $M$ that covers $a_1$, thus $a_2g_2$ can be qualified as the edge $a_1c_1$ in our notation. Using that case $vi)$ is applicable.

The last possibility is the one in which $a_1=a_2$ and no augmenting set is inserted into $M$.
Assume now that this is the case, thus the situation is as follows: $g_1y_1$ and $g_2y_2$ are in $M$, $a_1g_1=\mbox{shadow-edge}(g_1y_1,g_1)$ and $a_2g_2=\mbox{shadow-edge}(g_2y_2,g_2)$. $g_1y_1$ took over a $f_{A^\prime,k}(g_1)$-fraction of the charges from $a_1g_1$ when replacing it, $g_2y_2$ took over a $f_{A^{\prime\prime},k}(g_2)$-fraction of the charges from $a_2g_2$. Since $a_1=a_2$ it is also $c_1=c_2$.

Let $f_{A^\prime,k}(g_1)\cdot w(a_1g_1)\ge f_{A^{\prime\prime},k}(g_2)\cdot w(a_2g_2)$. It suffices to consider $y_1y_2$ as an optimal edge since otherwise no charge must be assigned if $y_1y_2$ is prevented and the properties of the lemma hold further on. 

Prior the arrival of $y_1y_2$, $coe(g_1y_1,y_1)=coe(g_2y_2,y_2)=0$, thus $a_1g_1$ and $a_2g_2$ are both prepared and fingering $a_1$. If $coe(a_2g_2,a_2)\curvearrowright ac(g_2y_2)=f_{A^{\prime\prime},k}(g_2)\cdot w(a_2g_2)+X$ for $X>0$, $a_2g_2$ is overloaded, thus $coe(a_1g_1,a_1)\curvearrowright ac(g_1y_1)\le f_{A^\prime,k}(g_1)\cdot w(a_1g_1)$ since $a_1g_1$ cannot be overloaded as well. $X$ cannot be greater than $(k-1)\cdot f_{A^{\prime\prime},k}(g_2)\cdot w(a_2g_2)$, therefore we can transfer a charge of weight $X$ from $coe(a_2g_2,a_2)\curvearrowright ac(g_2y_2)$ to $coe(a_1g_1,a_1)\curvearrowright ac(g_1y_1)$, $a_1g_1$ might get overloaded, $a_2g_2$ is not overloaded anymore.

After this transfer of charge, or if no transfer was necessary because $X\le 0$, we have $coe(a_2g_2,a_2)\curvearrowright ac(g_2y_2)\le f_{A^{\prime\prime},k}(g_2)\cdot w(a_2g_2)$. Thus $coe(g_2y_2,y_2)$ can take a charge of $k\cdot w(g_2y_2)$ without violating the properties of the lemma since in that case $coe(g_2y_2,y_2)$, $coe(a_2g_2,a_2)\curvearrowright ac(g_2y_2)$ and $ac(a_2g_2)\curvearrowright ac(g_2y_2)$ still satisfy an inequality corresponding to (1). If no augmenting set is inserted into $M$, $w(y_1y_2)\le\min\{k\cdot(w(g_1y_1)+w(g_2y_2)),\,k\cdot(w(g_1y_1)+w(a_1c_1))-w(a_1g_1)+k\cdot w(g_2y_2)\}$. Therefore the partial weight of $y_1y_2$ that $g_1y_1$ has to take as charge for preventing $y_1y_2$ satisfies the properties (A) or (B).

\ \\
We showed that the properties a)-d) of the lemma hold when $y_1y_2$ replaces and prevents edges. In the very same way the validity of the properties can be shown for the edges $a_1g_1$ and/or $a_2g_2$ that are possibly taken into $M$ at the same time as $y_1y_2$.\qed

\noindent Using Lemma \ref{Messy} we can prove our main theorem.
\\

\noindent{\it Proof of Theorem \ref{ApproxRatio}: } Let $M$ be the final $M_m$. $w(M^*)=w(M^*\cap M)+w(M^*\setminus M)$. Because for an edge $xy\in M^*\cap M$ we have $coe(xy,x)=coe(xy,y)=0$, we can write 
$$w(M^*\setminus M)\;\;\le\;\;\sum_{xy\in M^*\cap M}\frac{k}{k-1}\cdot w(xy)+\sum_{uv\in M\setminus M^*}T(uv)$$ That results in $w(M^*)\le\left(k+\frac{k}{k-1}+\frac{k^3-k+1}{k^2}\right)\cdot w(M)$.\qed

\noindent The term describing the approximation ratio of our algorithm reaches its minimum for $k$ being around 1.717, that yields a ratio of 5.585. It is easy to see that the algorithm does not exceed the space restrictions of the semi-streaming model: It needs to memorize the edges of $M$, for each of those at most two shadow-edges, thus it suffices to store a linear number of edges. The time required to handle a single input edge is determined by the size of $S$. Since $S$ is of constant size, a single run of the while loop, including the enumeration and comparison of all possible augmenting sets of $S$, can be done in constant time. Therefore the algorithm needs a per-edge processing time of ${\cal O}(1)$. As well as the single pass over the input this is optimal.


\subsection*{4 Conclusion}
We presented a semi-streaming algorithm calculating a weighted matching in a graph $G$. Our algorithm achieves an approximation ratio of 5.585 and therefore surpasses all previous algorithms for the maximum weighted matching problem in the semi-streaming model. In addition to the edges of an actual matching $M$ the algorithm memorizes some more edges of $G$, the so called shadow-edges. For each input edge $e$, the subgraph $S$ made up of $e$ and of shadow-edges and edges of $M$ in the vicinity of $e$ is examined. If a certain gain in the weight of $M$ can be made, matching and non-matching edges in $S$ are exchanged. 

The subgraph $S$ investigated by our algorithm for each input edge consists of at most seven edges. It is reasonable to assume that by examining bigger subgraphs the approximation ratio can be enhanced further. Therefore we believe that extending our approach will lead to improved semi-streaming algorithms computing a weighted matching.

\end{document}